\documentclass[twocolumn,showpacs,aps]{revtex4}
\usepackage[dvips]{graphicx}
\newcommand{\beq}{\begin{equation}} 
\newcommand{\beqar}{\begin{eqnarray}} 
\newcommand{\eeq}[1]{\label{#1} \end{equation}} 
\newcommand{\eeqar}[1]{\label{#1} \end{eqnarray}} 
\begin{document}
\title{A study on a self-organized criticality in a dynamical
many-body system}
\author{Akira~Iwamoto}
\affiliation{Japan Atomic Energy Research Institute, Tokai, Ibaraki, 319-1195 Japan}
\author{Shinpei Chikazumi}
\affiliation{Department of Theoretical Studies, Institute for Molecular Science, 
Okazaki, 444-8585 Japan}

\begin{abstract}

A novel mechanism for the generation of self-organized criticality (SOC) 
is discussed in terms of
the coupled-vibration model where the
total system is forced under the uniform expansion of the Hubble type. 
This system shows a robust SOC behavior while the maximum size of the
fluctuation, number of correlated particles in it and the temporal size of the system
evolve as a function of time.  

\end{abstract}
\maketitle

Observation of the ubiquitous manifestation of fractal structure 
in nature is one of the most fascinating findings in non-linear
science \cite{mandelbrot83}.  One direction for the understanding of the
origin of its manifestation was offered in \cite{btw87,btw88} 
with the concept of self-organized criticality (SOC). It was argued 
there that the scaling properties  of the attractor is insensitive to the 
parameter of the model, which shows a robustness of the criticality. 
Since then, many models of cellular-automata were studied 
\cite{jensen98,turcotte99}, where another characteristic feature is that
the external driving of the system is much slower than the internal
relaxation processes.   

The separation of external and internal
time-scales, however, is not always a guiding principle for the pattern formation.
Turbulent flow, for example, seems to have more dynamical origin. 
In this respect, we will seek a SOC state where the external driving of
the system has a comparable time scale to that of the internal motion. The 
disappearance of the characteristic length scale and the robustness are used 
to characterize SOC.  As a typical example which shows this feature, we propose a
homogeneously expanding lattice model which was motivated by our previous  
study \cite{ci04}.  We define 
a one-dimensional version of the previous model\cite{ci04} and make an anatomy of 
SOC formation when the many-particle system
is put in a mechanically unstable state. 

Our model is based on an expanding one-dimensional coupled-vibration model.  The chain is
composed of $N_0+2$ particles of equal mass $m$, coupled with the nearest
neighbor particles by a finite-depth potential.  This chain is forced to obey
a uniform expansion of the Hubble-type \cite{ci04}. Physically important 
feature is that as the lattice spacing increases, each
particle at the lattice point becomes unstable against the motion 
to right or to left, forming a cluster.  Although the following discussion 
will be applied to rather general class of interactions, we take Lennard-Jones (L-J)
potential as a prototype \cite{ci04}.

\begin{figure}
\begin{center}
\includegraphics[width=7cm,clip]{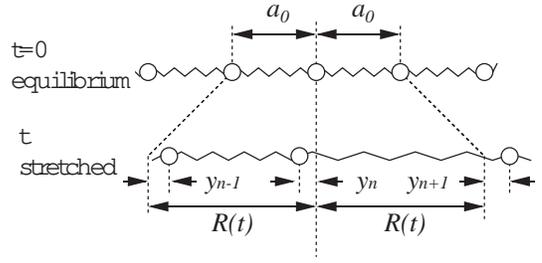}
\end{center}
\caption{Sketch of the extended chain used.  At equilibrium, random and
small fluctuations are assumed.}
\end{figure}

We analyze a 
motion of mono-particle one-dimensional lattice with a lattice 
spacing $R$ larger than its equilibrium separation $a_0$ as is given schematically 
in FIG.1.  The expansion of the
lattice is described with the relation, 
\beq
R(t) = a_0 (1+ht)
\eeq{eq:expansion}
where the Hubble constant $h$ is introduced.  This condition is 
satisfied by putting the initial velocities to each particle together with the
corresponding velocities to the boundaries to which $0-th$ and $(N_0+1)th$
particles are held fixed.  The equation of motion of this system
is written as
\beq
    m\frac{d^2 y_n}{dt^2 } = V'(R + y_{n+1} - y_n ) - V' (R + y_n - y_{n-1} ),
\eeq{eq:eom1}
where $y_n$ stands for a displacement of n-th lattice particle from its lattice point
and $V'$ is the derivative of the inter-particle potential. Suffix n runs from
1 to $N_0$ in this equation.  We assume this potential 
$V$ is given by the Taylor polynomial
expansion around $R$ for small displacement $r$,  
\beq
    V'(R+r) \approx V'(R) + V''(R)r  .
\eeq{eq:Taylor}    
Inserting this form to (\ref{eq:eom1}), we obtain
\beq
    m\frac{d^2 y_n}{dt^2 } =  \kappa ( 2y_n - y_{n+1} -y_{n-1} ) ,
\eeq{eq:eom3}
where we put $\kappa$ for $-V''(R)$.
It should be noted here that the $\kappa$ is always
negative when we deal with the stationary oscillation where $R=a_0$.  On the
other hand, for L-J potential, it becomes positive when $R \geq 1.11a_0$.  In the 
following discussion, we mainly treat this situation and thus we assume $\kappa$
is positive.

To solve (\ref{eq:eom3}),  we 
assume the following standard form of the solution,  
\beq
  y_n = a\exp ( ik n \mp \omega t ) .
\eeq{eq:trial}
Inserting this form to eq.(\ref{eq:eom3}), we get,
\beq
\omega ^2 = \frac{4\kappa }{m} \sin^2 \frac{k}{2}.
\eeq{eq:disp}
This dispersion relation has the same form as for the normal-mode analysis for
the equilibrium oscillation except for the fact that the motion of the lattice
particle  $y_n$ is not oscillatory but divergent as is clear from eq.(\ref{eq:trial}).
When we calculate the second derivative
of $y_n$ in (\ref{eq:eom3}), the following derivative appears,
\beq
d(\omega t)/dt  = \omega + d\omega/dt \times t.
\eeq{eq:check}
For a non-singular form of $\omega$, the second term in the rhs is smaller than the
first term for sufficient small value of t.  Therefore, we can justify the dispersion
relation (\ref{eq:disp}) under the condition $y_n \ll R(t)$ and for short-time regime.

The solution of the equation (\ref{eq:eom3}) for small displacement and for 
small t, under the boundary condition,
\beq
y_0 = 0 = y_{N_0+1}
\eeq{eq:bc}
is written as
\beq
y_n = a \exp (\mp \omega t) \sin (kn),
\eeq{eq:sol1}
where $a$ is a constant.
In this form, the boundary condition for n=0 is automatically satisfied and for n=$N_0$+1,
we obtain a relation,
\beq
k = \frac{\pi \nu}{N_0+1}.\,\,\,\, (\nu = 1,2,...N_0) 
\eeq{eq:bc2}
For convenience, we rewrite (\ref{eq:bc2}) and (\ref{eq:disp}) in the form,
\begin{eqnarray}
&k_{\nu} = \frac{\pi \nu}{N_0+1},& \label{eq:arelation}\\
&\omega_\nu ^2 = \frac{4\kappa }{m} \sin^2 \frac{k_{\nu}}{2}.&\,\,\,\, (\nu = 1,2,...N_0) 
\label{eq:brelation}
\end{eqnarray}

The general solution is written down in the form
\begin{eqnarray}
&y_n = \sum A_{n\nu}Q_\nu, \label{eq:sol1-1} \\
&A_{n\nu} = \sqrt{\frac{2}{N_0+1}} \sin \frac{\pi n\nu}{N_0+1}, \label{eq:sol1-2} \\
&Q_\nu = a_\nu [\exp(\omega _\nu t) + b_\nu \exp(-\omega _\nu t)]/2, \label{eq:sol1-3}
\end{eqnarray}
where $a_\nu$ and $b_\nu$ are free parameters that fit the 2$N_0$ initial
conditions for $N_0$ particles.

In the $\nu$-th normal-mode for the longitudinal motion (identify $nR$ as 
x-coordinate), the following
number of particles $n_\nu$ are compressed or stretched as a whole,
\beq
n_\nu = \frac{(N_0 + 1)}{\nu} = \frac{\lambda _\nu (t)}{2R(t)},
\eeq{eq:corrpar}
where $\lambda _\nu$ is a wavelength for the $\nu$-th mode expressed as 
$(2L(t))/\nu$ by using the total length of the chain $L(t)$ at time t.  Therefore, 
the correlation of $n_\nu$ particle is produced by the $\nu$-th normal
mode.  
The growth of the $\nu$-th normal mode is governed by (\ref{eq:sol1-3}) 
and is exponentially divergent.
The time ordering of the growth is known from eqs.(\ref{eq:arelation})
and (\ref{eq:brelation}).  For larger $\nu$, the $k_\nu$ value becomes larger 
, which leads to larger
$\omega_\nu$ value.  It means that the exponential
growth of the fluctuation occurs stepwise, from the highest frequency to the
lower frequency, in other word, from the shortest wavelength to the larger
wavelength. 
When we start from the random small fluctuations of
the particles' position or velocities or both, it corresponds to the fact 
that $a_\nu$  in (\ref{eq:sol1-3}) stands for a white spectrum.

Above discussion is applicable only for small fluctuation and for short-time 
behavior. In other word, it is a linearized discussion because we assumed no
coupling between various normal modes.   For the analysis of SOC, however, 
we have to go into finite-time and nonlinear regime.  We first postulate 
the following three hypotheses:
1) Even at finite-time, the formation of cluster starts from
small particle number to large particle number.  
2) At the time when $n_\nu$ correlation just starts, its spatial size 
$\delta_c (t)$
is characterized by the relation
\beq
 \delta_c (t) \equiv \lambda_\nu (t) / 2 = n_\nu (t) \times R(t)
\eeq{eq:basic1}
3) The spatial size of the fluctuation for $n_\nu$ particles is approximately conserved even at
later time. 

The hypothesis 1) and 2) are taken from the linear regime 
from (\ref{eq:corrpar}).  
The hypothesis 3) means
that the size of the clusters are frozen at the time of its formation, without
stretching or shrinking at later time.  It means that at
the start of the $n_\nu$ correlation and later on, $n_\nu$ particles
are trap in the mutual attractive interaction bloc, free from the uniform expansion 
to which $n_\nu$ particles followed before.  Since we assume no energy
dissipation, the time-averaged spatial size of the bloc is conserved. It should be noted that when
$n_\nu$ correlation starts, correlations of smaller particles are preserved in
the $n_\nu$ correlation, forming a nesting-boxes structure.

Next task is to determine the dependence of the $\delta_c (t)$
and $n_\nu (t)$ on $R(t)$ which satisfies (\ref{eq:basic1}). We postulate the 
following relation to generate the solution of (\ref{eq:basic1}).
\beq
\frac{dn_\nu (t)}{d\delta_c (t)} = \gamma / R(t)
\eeq{eq:basic2}
where $\gamma$ stands for a non-dimensional constant.  The form of lhs was adopted because we
study a growth process of $n_\nu$ and $\delta_c (t)$. The form of rhs is assumed because the lhs has
the dimension of [length]$^{-1}$ and we know that the system has a lattice unit parameter $R(t)$ 
which has the dimension of [length].  The parameter $\gamma$
is a function of the Hubble constant $h$ in such a way that $\gamma \rightarrow 0$ for large
value of $h$ and $\gamma \rightarrow 1 $ for small value of $h$.  The former
condition corresponds to an enlarged copy of the system discussed in \cite{ci04} 
and the latter to a system where there happens one break of the chain and the left and the right 
matters move to the left and to the right without changing their sizes. 
Thus the assumption of (\ref{eq:basic2}) is that
lhs is proportional to the bulk ratio  $R(t)$, cf.(\ref{eq:basic1}), times $\gamma$, which stands for
the amount of tension due to the competition of expanding motion and 
the internal attraction.  The condition that this $\gamma$ is
a constant is the condition of realizing the SOC, because we will see in FIG.3 that 
the relation (\ref{eq:basic2})
is found to be well satisfied in the numerical calculations.   

Using (\ref{eq:basic1}), the solution of (\ref{eq:basic2})
is written as
\beq
n_\nu (t) = const \times \delta_c (t)^\gamma.
\eeq{eq:sol2}
This $n_\nu$ is directly related to the box-counting number $N(t)$ 
for the size of $\delta_c$, as is given by the relation,
\beq
N(t) = N_0/n_\nu(t) = N_0 \delta_c (t)^{-\gamma}/const ,
\eeq{eq:sol3}
which is clear from (\ref{eq:basic1}). The relation (\ref{eq:sol2}) means 
that the correlation length is expressed as a power of the distance, a typical feature 
of phase transition point. Here, $n_\nu$ is considered as distance measured
in unit of lattice spacing $R(t)$. Since $\gamma$ depends on the
expansion speed, the criticality is not universal here but SOC is well satisfied because 
the system without fail goes into the critical state that  satisfies the power-law relation (\ref{eq:sol2}).
The fact that all the complexity of the non-linear effect is simply written in a form 
(\ref{eq:basic2}) is astonishing.

\begin{figure}
\begin{center}
\includegraphics[width=7cm,clip]{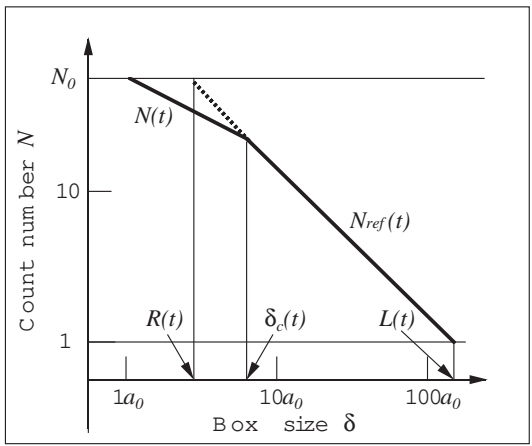}
\end{center}
\caption{Relation of $\delta_c(t)$, $N(t)$ and $R(t)$ in the graph of log-log sale 
for the box size $\delta$ and count number $N$.  Cross-over effect and the maximum fluctuation are 
shown.}
\end{figure}

Inserting (\ref{eq:sol2}) into (\ref{eq:basic1}), we can get the explicit expression of $\delta_c(t)$ 
in terms of $R(t)$.  We can also obtain $\delta_c(t)$ in terms of the relation among
$\delta_c (t)$, $N(t)$ and $R(t)$ with the help of FIG.2 drawn in log-log scale 
with the abscissa corresponding to box size $\delta$ and the ordinate, to the
count number $N$.  
In this figure, 
the trace of (\ref{eq:sol3}) is shown by the solid line $N(t)$ up to a specific
time t.
For a fixed time, the maximum size of the fluctuation is 
$\delta_c (t)$, beyond which the box counting number should follow the one-dimensional
line $N_{ref}(t)$ shown in the figure. That is, we define $\delta _c (t) $
and $N(t) = N_0/n_\nu (t)$ as x- and y-coordinates of this cross point. The line $N_{ref}(t) $
should pass through the count of
unity when the box size $\delta$ coincides with the total chain length L,
\beq
\log{N_{ref}(t)} = -\beta (\log{\delta} - \log{L(t)}),
\eeq{eq:ref}
where $\beta$ stands for 1 in our case and 3 in the 3-dimensional case.

The value of $\delta_c (t)$ is the x-coordinate of the cross point of
the line (\ref{eq:sol3}) and the line (\ref{eq:ref}) which is expressed as
\beq
\delta_c (t)^{\beta - \gamma}  = const \times  R(t)^{\beta},
\eeq{eq:deltac}
which has the same form as the short time behaviour seen in dynamical scaling \cite{vf84}.
 For general spatial dimension, relation (\ref{eq:basic1}) is replaced by
\beq
 \delta_c (t)^\beta = n_\nu (t) \times R(t)^\beta,
\eeq{eq:basic3}
where $\beta$=3 applies to our previous calculation \cite{ci04}. 

The third hypothesis we postulated is a key to get the fractal growth in SOC state
and the normal fractal dimension calculation for a fixed system. 
Because of this hypothesis, the fractal structure formed in SOC is frozen out.  
The fact that the growth of
the maximum fluctuation in our model is closely related to a cross-over effect, is a typical feature
of our dynamical SOC state.

\begin{figure*}
\begin{center}
\includegraphics[width=14cm,clip]{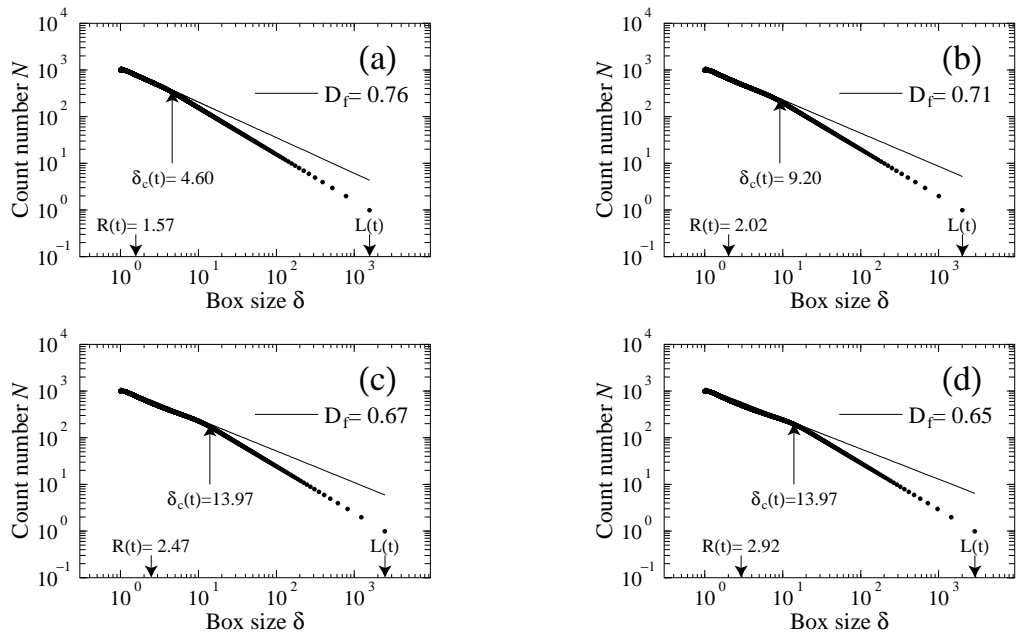}
\end{center}
\caption{Numerical simulation of the box-counting for the system for $h$=0.1, $N_0$=1000 averaged over
10 events for fixed boundary condition and for nearest neighbor interaction.  Each panel corresponds
to the time sequence of the expansion of the system.}
\end{figure*}
 
Now we are ready to check our three hypotheses and the postulation (\ref{eq:basic2}) with a help 
of the numerical simulation for linear chain interacting with L-J potential The unit of length 
$\sigma$ is the distance at which the potential
changes sign and the energy unit $\epsilon$ is the depth of the potential.  The time unit $t_0$ is 
composed as $t_0 = \sqrt{\sigma^2m/48\epsilon}$. Before physical discussion,
we first check the dependence of the 
results on the number of particles $N_0$.  We performed 
the box-counting method with 100 particles and 1,000 particles. The results hardly change
when we averaged over 10 independent trials for the different initial states. 
For boundary conditions, we used a fixed boundary condition (\ref{eq:bc})
in our formalism.   We checked the periodic boundary condition 
to simulate the infinite system \cite{ci04}.  We also checked the results when we include only
the nearest neighbor interaction or allow the next neighbor one.  We found that such changes
altogether cause at most 1\% change of the fractal dimension value.

In FIG.3, we show the box counting plot for the system for $h$=0.1, $N_0$=1000 averaged over
10 events for fixed boundary condition and for nearest neighbor interaction. The value of
the lattice spacing is shown in these figures by $R(t)$ in unit of the equilibrium lattice 
spacing.  As we see in this figure, the fractal structure is clearly seen up to the cross-over
point, which extends as $R(t)$ increases. 
The linearity of the slope, which is closely related to the assumption of constant $\gamma$ 
in eq.(\ref{eq:basic2}) is well realized in this figure.  
The FIG.3(d) corresponds to $R(t)$=2.92, which exceeds the cut-off length 2.5 of our
Lennard-Jones potential \cite{ci04}.  In this situation, the formation of a cluster in our
formalism is not physically  applicable and therefore, the $\delta_c$ value of FIG.3(c), which is
just before $R(t)$ exceeds the cutoff length is shown. Actually, the behavior beyond this 
$\delta_c$ value starts to wave from the linear line. Another important tendency is the
gradual decrease of the slope parameter as time proceeds.  The change of this value is interpreted
as due to the decrease of the stiffness parameter $\kappa$ as $R(t)$ increases.  This causes
the decrease of the clustering force compared with the expansion induced by $h$, which leads to a
smaller fractal dimension $\gamma$. 

The same degree of good fitting with two linear lines is obtained for h=0.2,0.3 and 0.4.  
As is expected, the fractal dimension
becomes smaller, the larger the $h$ values.  For example, the slope parameters (fractal dimensions)
corresponding to FIG.3(c) ($R(t)=2.47$) are 0.67, 0.61, 0.56, 0.51 for $h$=0.1, 0.2, 0.3 and 0.4. 
Another important factor which determines the slope is the strength of the initial fluctuation.
We adopted the initial conditions that $N_0$
particles are initially at rest and are located around the lattice site randomly within 
$\pm p\%$ of the original lattice spacing.  This condition, when we
remember eq.(\ref{eq:sol1-3}), corresponds to $b_\nu = 1$  and the strength of 
random variable $a_\nu$ is proportional to $p$.  The results shown in FIG.3 corresponds to $p$=5.  
The fractal dimension for $p$=5 $\sim 10$ is rather stable, changes less than 3 \% like three
dimensional calculation in \cite{ci04}. However, 
when we reduce $p$ to smaller values, the fractal dimension becomes smaller, approaching to
zero for small fluctuation limit.  It is because 
$a_\nu$ value in (\ref{eq:sol1-3}) becomes smaller for smaller initial fluctuation and the start of $n_\nu$
correlation is delayed with respect to time, which means the decrease of the slope of the
fractal line. The numerical calculations show essentially the same results as in the 
3-dimensional case \cite{ci04}, which suggests that this kind of SOC is common to any dimension.

In conclusion, our coupled linear-chain model with finite-range interaction, when forced under
the uniform expansion of the Hubble type, leads to a robust SOC state and shows a 
fractal structure and a cross-over effect. 
The slope (box-counting fractal dimension) depends on the expansion speed and the strength of
the initial fluctuations.  This SOC state in the dynamical 
many-particle system differs from the standard SOC in two respects, 1) relative time scales of
external and internal motions, 2) existence of a clear cross-over effect, that has a possibility to 
open a new window towards our understanding of fractal structure in nature.

We thank Dr. T. Yokota for useful discussions.
 

\end{document}